# Survey on Multi-Document Summarization: Systematic Literature Review

**Uswa Ihsan[1], Humaira Ashraf[1], NZ Jhanjhi[2]**

1. **Department of computer science and software engineering, International Islamic University Islamabad.**
2. **School of Computer Science, SCS, Taylors University, Subang Jaya, Malaysia**

**Emails: uswaihsan.mscs1075@iiu.edu.pk, humaira.ashraf@iiu.edu.my, noorzaman.jhanjhi@taylors.edu.my**

## Abstract

In this era of information technology, abundant information is available on the internet in the form of web pages and documents on any given topic. Finding the most relevant and informative content out of these huge number of documents, without spending several hours of reading has become a very challenging task. Various methods of multi-document summarization have been developed to overcome this problem. The multi-document summarization methods try to produce high-quality summaries of documents with low redundancy. This study conducts a systematic literature review of existing methods for multi-document summarization methods and provides an in-depth analysis of performance achieved by these methods. The findings of the study show that more effective methods are still required for getting higher accuracy of these methods. The study also identifies some open challenges that can gain the attention of future researchers of this domain.

## 1. Introduction

Now days, a huge amount of data is available on the internet in the form of web pages and documents on any given topic. It is difficult for users to extract useful information from the collection of documents. The technique of producing a compressed version of a given source text that provides valuable information for a specific user and task is known as Automatic text summarization [1].

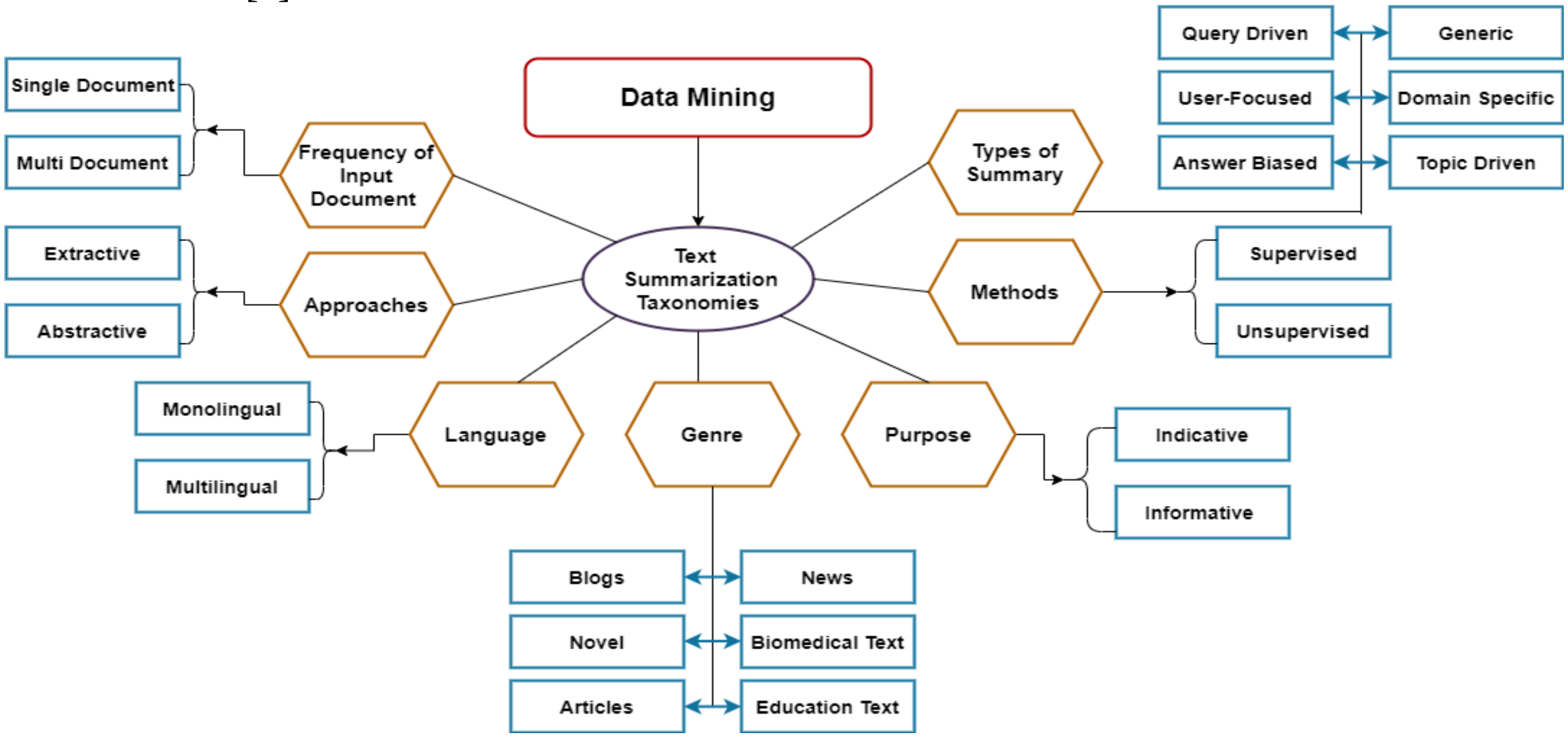

*Figure 1 Data Mining Techniques [2]*

In figure 1, the Data mining technique is discussed. In-text summarization, frequency of input document is the single document and multi-document, approaches used for summarization are Extractive and Abstractive, methods used for summarization are supervised/unsupervised and different types of the summary are query-driven, Topic driven, user-focused, generic and domain-specific.

Multi-document summarization is distributed into two types abstractive and extractive. Extractive summarization generated the summary by adding the most relevant information in the sentence of the summary. It is more feasible and used for sentence ranking[18] [57][21]. The Extractive summarization system can be distributed into two methods supervised and unsupervised. A supervised system gets qualified from labeled data to pick useful content from documents. An Unsupervised system is very Well-known because it does not need manually labeled data for generating a summary from documents[58]. Automatic extractive summarization can be done using three techniques: preprocessing, phrase grading, and clause selection[20]. The extractive multi-document summarization intends to extract useful information from the collection of documents, maintain the relevant data, and reduce the redundancy from it[17]. Abstractive summarization understands the original text from the document and then creates new sentences from it [18] [57] [21].

Based on previous research in this field there are the following types of summaries: Generic-based summarization extracts the main context of the document without any rest of specific information. Doubt-based summarization generates a precise summary of a particular set of the document and a given query [18] [3] [59]. Term based method efficiently gives the result based on term weight calculation [18].It is further divided into centroid and graph-based methods that belong to extractive summarization. In the centroid-based method, a clustering algorithm is used for the selection of the sentence representation from the different clusters whereas the graph-based method selects a sentence based on voting from their neighbor using ideas [59].

The main challenge in multi-document summarization is redundancy. In the existing literature, researchers proposed a lot of methods to improve redundancy issues. Patel and Shah et al [2], create a method that provides good content coverage with information diversity and improved the significant performance as compare to other summaries. Mani and Verma et al [3], develops a method for the least rebuilding error while creating summaries, and improves the performance of the system. Hafeez and khan et al [4], proposed a method to improve the performance as compared to other systems and create an effective and human-readable summary. Li and peng et al [5], develop a method to enhance the feature and readability of the created summary and also achieve higher performance and reduce the redundancy. Zhao and Liu et al [7], develop a method that is used to improve the performance of the system and generate more readable and natural summaries. Peyrard and Eckle-Kohler [49], It combines surface level information with frequency features to represent sentences (location, lengths, and overlap with title). It includes TF*IDF term weighting: Calculate the sum of the occurrence of the bi-grams in the sentence, in addition the sum of the document frequency of the terms and bi-grams in the sentence. Dong et al [50], Sentence selection is represented as a sequential sampling without replacement process, with the sampling likelihood determined by the normalized scores of the remaining sentences. Sentence scores are often not changed depending on the current partial summary of previously picked sentences in these systems, demonstrating a lack of information about the extraction history. However, the method used by Mao and Yang et al [21] is affected by inaccurate similarity results toward vectors' magnitudes. Cagliero and Garza et al [22], the beam search algorithm is used which is complex because multiple transmittances of broad beams raise the beam arrangement delay, and assigning more power to transmit broad beams and less power to transmit restricted beams requires a broad vigorous range for power amplifiers. Noh and song et al [23], a latent semantic analysis algorithm is used which correlated with only single document terms. The mostly dataset used in above papers are DUC 2001 DUC 2002, DUC 2004, DUC 2006, DUC 2007, TAC08, TAC11, and MultiLing13 etc.

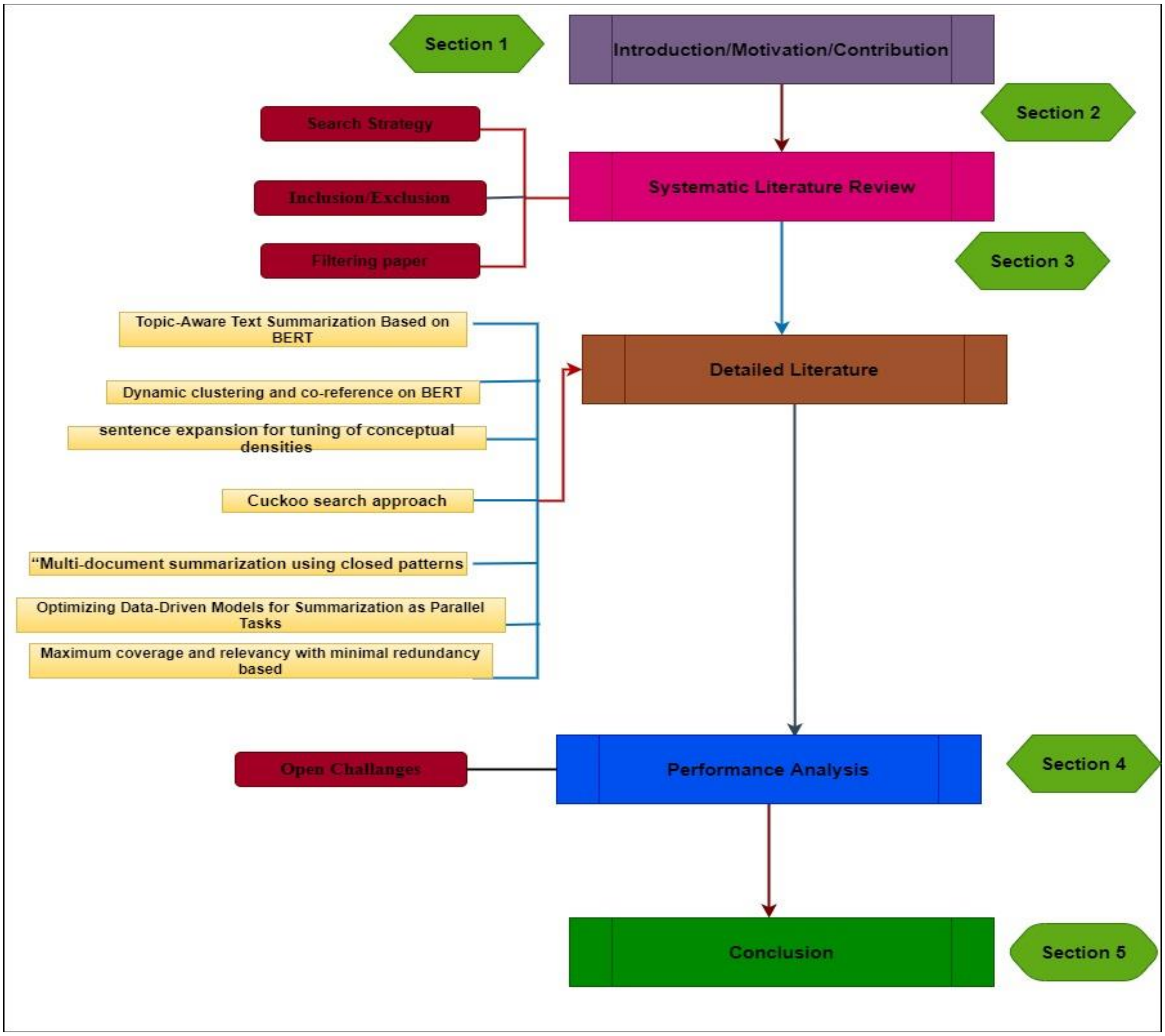

*Figure 2. Organization of Survey*

The motivation of this research paper is the collection and in-depth study of paper which addresses the issue of redundancy. In this paper after an extensive search with strings on various databases collected 140 papers then screening the paper through title-based and abstract-based evaluation to assess the quality of whether the research is suitable or not.

The Systematic literature review (SLR) concludes although the researcher proposed many techniques for reducing redundancy. However, there is also a need to create some techniques that improve the accuracy with better algorithms and also create high-quality summaries.

In table 1, Existing surveys on multi-document summarization are discussed. This table shows the strength of this survey paper with already existing survey papers using the comparison method. Existing survey papers only presented limited features whereas this research paper overall handle all feature like literature scheme achievements, limitations, research gaps, future work, performance analysis etc.

| Year | Survey Topic | Methods | Enhancement in our paper |
|---|---|---|---|
| **2004** | A Survey on Multi-Document Summarization [30] | The paper handles three components of multi-document summarization.<br>1. Centroid-based summarization<br>2. Information fusion algorithm<br>3. Sentence compression by using two algorithms. | Paper [30] only considered 3 component of multi-document summarization whereas this research discusses and compare multiple components of multi-document summarization in detail. Also, compare the survey based on multiple objects. |
| **2017** | Text Summarization Techniques: A Brief Survey [31] | The paper only considers extractive approaches. | This survey paper considered both abstractive and extractive approaches of multi-document summarization and also discuss the comparison of techniques. The achievements and limitations of different papers are also discussed in our paper. |
| **2021** | Automatic text summarization: A comprehensive survey [53] | The survey paper presented approaches, procedures, methods, datasets, evaluation methods, and future work of automatic text summarization.<br>However, there is no performance analysis based on quality assessment. | This study provided a complete performance analysis, which included a critical analysis, a result comparison, and a gap analysis of all existing schemes. |
| **2021** | A Survey of Data Augmentation Approaches for NLP [54] | The paper only presented methodology of literature on data augmentation for NLP.However, this is not a systematic survey and the advantages and disadvantages of existing methods are not specified. The performance analysis of scheme is also not included. | Paper [54] only discuss the methodology of the existing scheme however this research presented techniques, limitation, achievement and dataset. This survey paper also discusses the performance analysis of scheme on the base of quality assessment. |
| **2022** | Compression of Deep Learning Models for Text: A Survey [55] | The paper handles only Deep learning models for NLP. However, this is not a systematic survey and the advantages and | Paper [55] only considered Deep learning models for text summarization where as our survey paper discussed both deep learning |
| | | disadvantages of existing methods are not specified. | model and other handmade model. This survey paper used SLR protocol. |

### *1.1 Motivation:*

The latest trend is the desire to learn about the specific field without dedicating months and hours to it. The need to meet this challenge motivates demanding work related to text summarization technology. Nowadays, no one has the time to read a huge number of documents from the internet and then extract the most relevant and useful information from them. In this time, multi-document summarization gives a more precise and comprehensive summary from the collection of documents into a single document and saves them time and efforts of the readers. Today, a lot of methods are already developed in this field for improving the performance of technique but existing work needed more accuracy.

### *1.2 Research Contributions*

In research contribution we achieve:

- Explore existing work to overcome the redundancy problems.
- This research is based on SLR (Systematic Literature Review) criteria.
- Research focuses on the comparison of the techniques that exist in literature and find the research gap between them.
- Critical analysis of existing literature is presented.
- The research gap in the field is analyzed and highlighted.

The next section is a systematic literature review, the third section is a performance analysis and the fourth section is a conclusion.

## 2. Systematic Literature Review

The main objective of SLR is to provide a comprehensive summary of the given literature relevant to a searched question. The strings developed by using the objective of all papers and then used 3 synonyms of all keywords and create all possible strings. 125 strings are created for the research question mention below. Search each string on three different databases and take the research paper randomly from it. In inclusion criteria, characteristics that the future subjects must have if they are to be included in the study whereas exclusion criteria are those characteristics that exclude the future subjects in the study. From the collection of the research paper, remove duplicate paper then title-based and abstract based screening is performed.

### 2.1 Protocol Development Search:

The protocol development search is use in systematic search for finding objectives, and identify gap analysis etc. for any specific research topic.

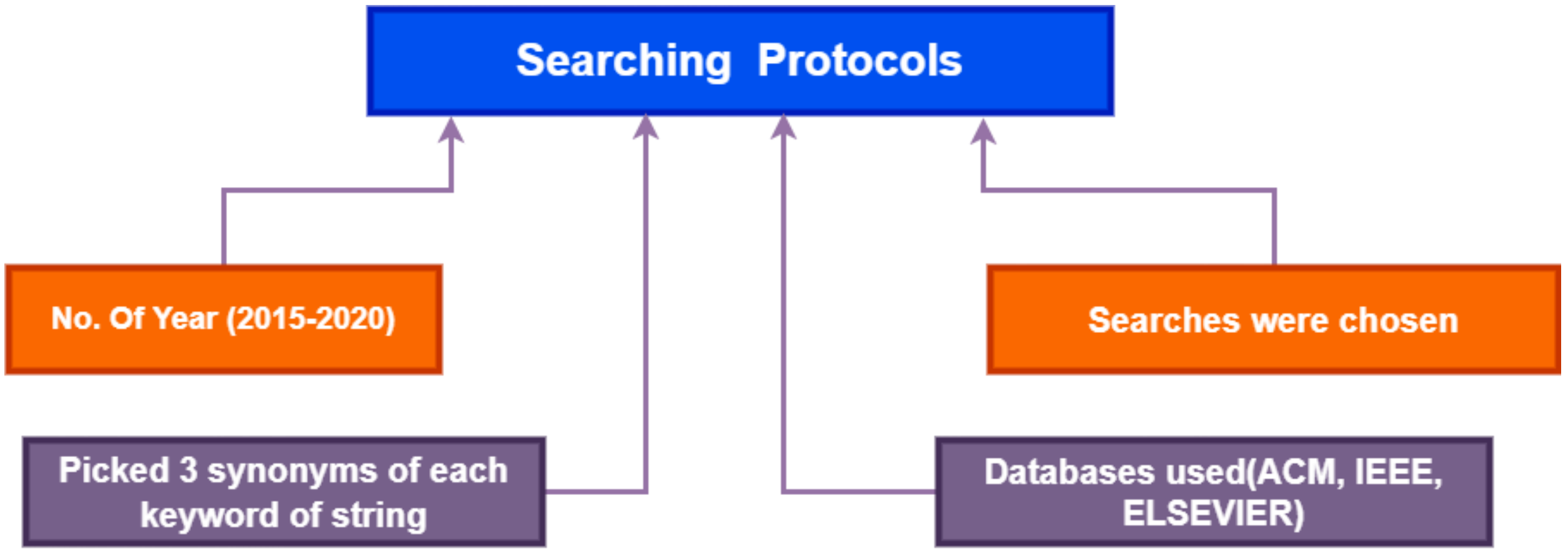

Figure 3: Protocol Development Search

In figure 3 we select number of year (2018-2020), picked 3 synonyms of each keyword of string and for searching used 3 database (ACM, IEEE, ELSERVIER) and then choose random searches against strings.

## 2.2 Search Strategy:

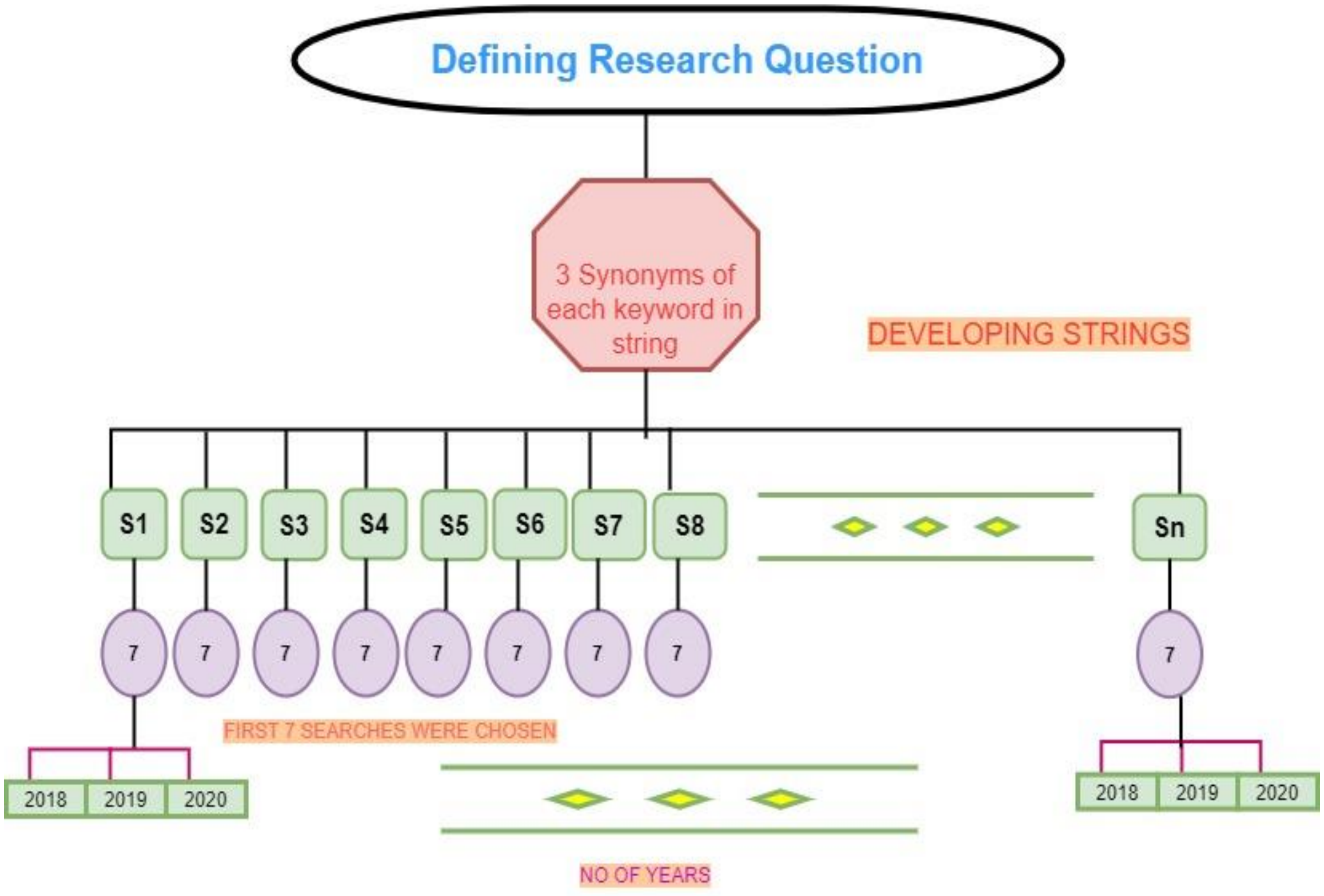

Figure 4: Search Strategy

In figure 4 search strategy discussed, in which firstly define a research question, use 3 synonyms of each keyword in string, and create all possible strings of 3 synonyms and search in three different databases**.**

## 2.3 Development of Strings:

The strings developed by using the objective of all papers and then used 3 synonyms of all keywords and create all possible strings. 125 strings are created for the research question mention below. Search each string on three different database and take research paper randomly from it

**Question of Research:** How to detect or minimize redundancy in multi document summarization.

## 2.4 Inclusion/Exclusion Criteria:

In inclusion criteria, Characteristic that the future subjects must have if they are to be included in the study where as Exclusion criteria are those Characteristics that exclude the future subjects in the study.

## 1. Primary Search:

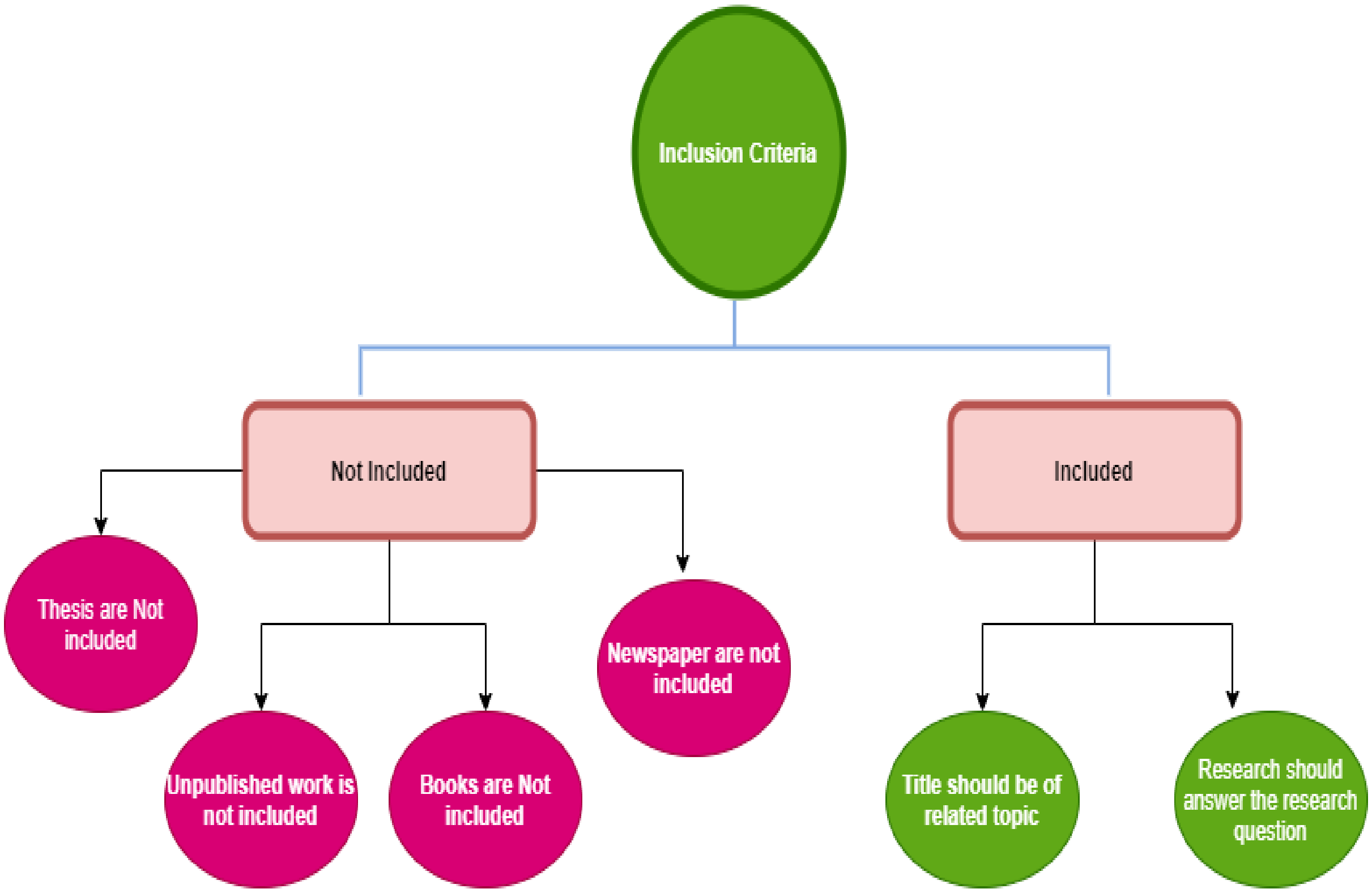

*Figure 5: Inclusion criteria*

In figure 5, inclusion criteria are discussed. There are two parts one is included and another is excluded (not included). The thesis, newspaper, books and unpublished work are not included in inclusion criteria and title based and research papers are included.

## 2.5 Duplicates Removal:

From the collection of Research paper, remove all the paper whose title names are the same because it creates duplication only to take one from them for further working.

## 1. 1st Step Title Base Filtering:

Title based filtering is discussed, read all papers title and then include papers that are relevant to the topic which is used for research and exclude papers that are irrelevant.

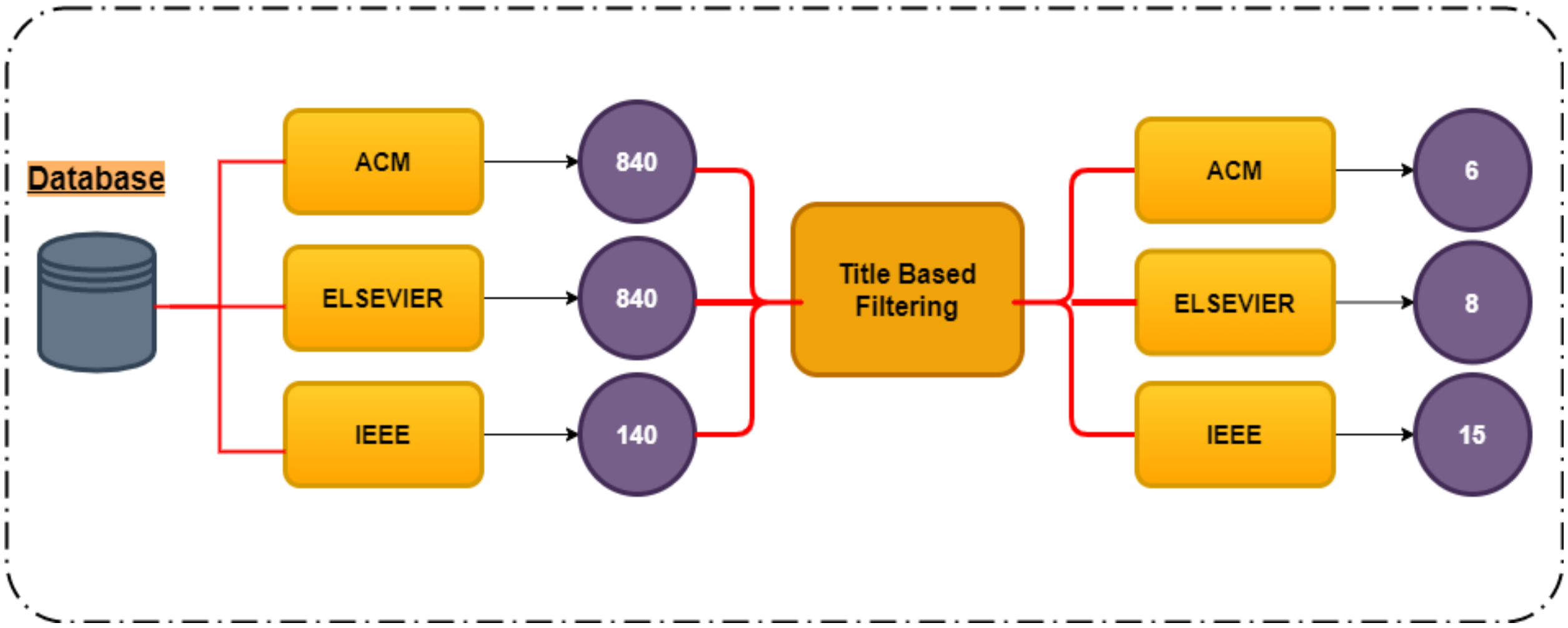

***Figure 6 Title Based Filtering***

In figure 6, from all three database collect research paper then apply title-based filtering on that research paper.

## 2. <u>2nd Step Abstract Based Filtering:</u>

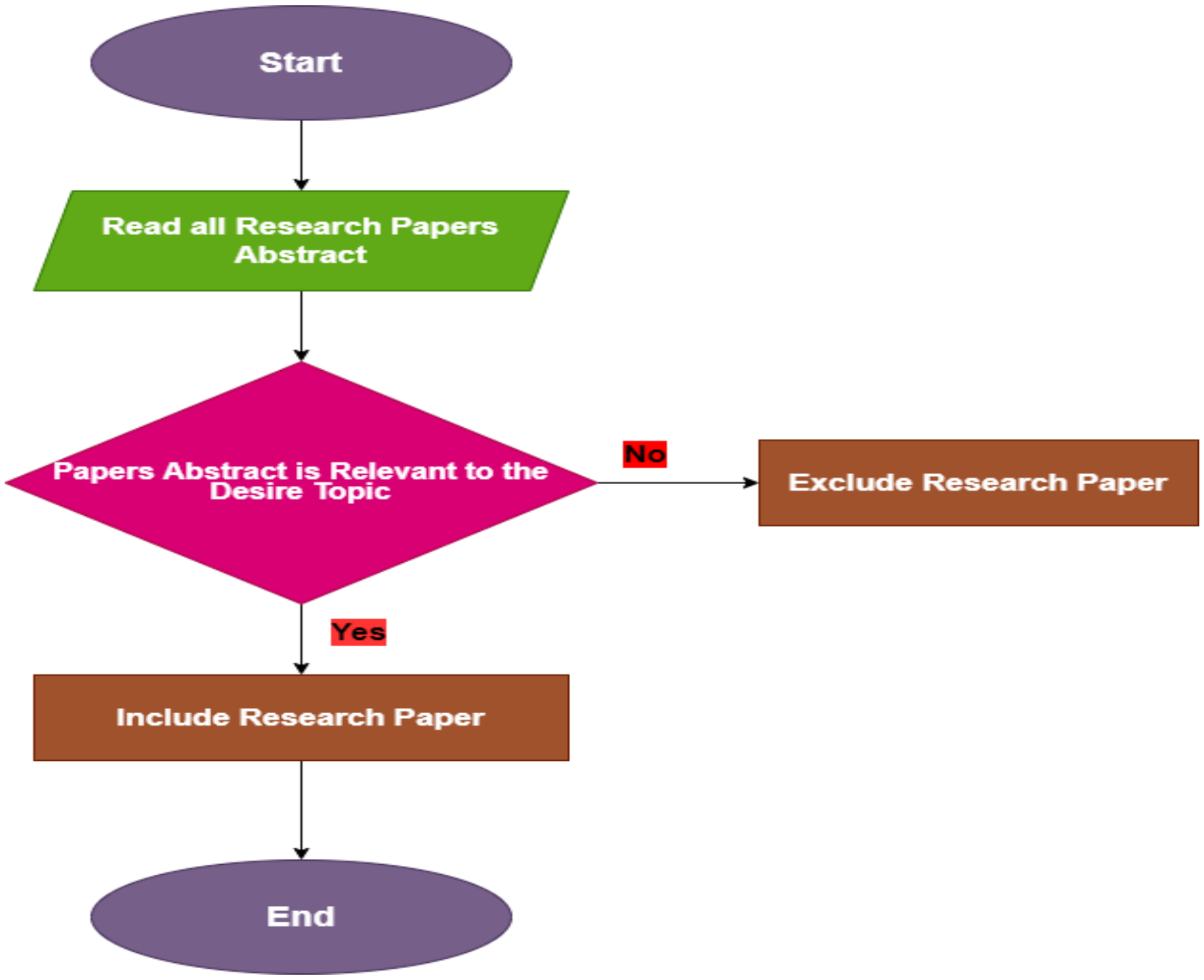

*Figure 7: Flow Chart of Abstract Based Filtering*

In figure 7, Abstract based filtering is discussed, read all research papers abstract carefully and then include papers that are relevant to the topic which is used for research and exclude papers that are irrelevant to the research topic.

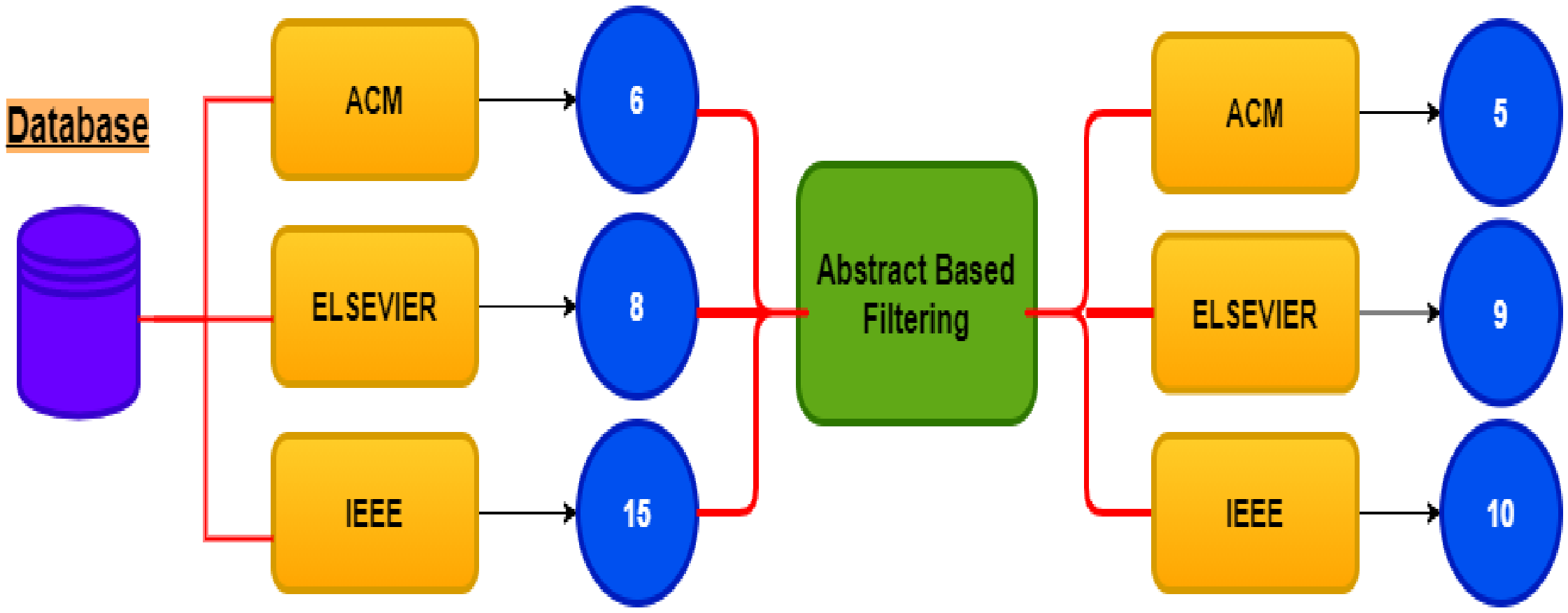

*Figure 8: Abstract Based Filtering*

In figure 8, after applying title-based filtering the paper left apply abstract based filtering on it also remove redundant paper.

## 2.6 Objective-Based Clustering:

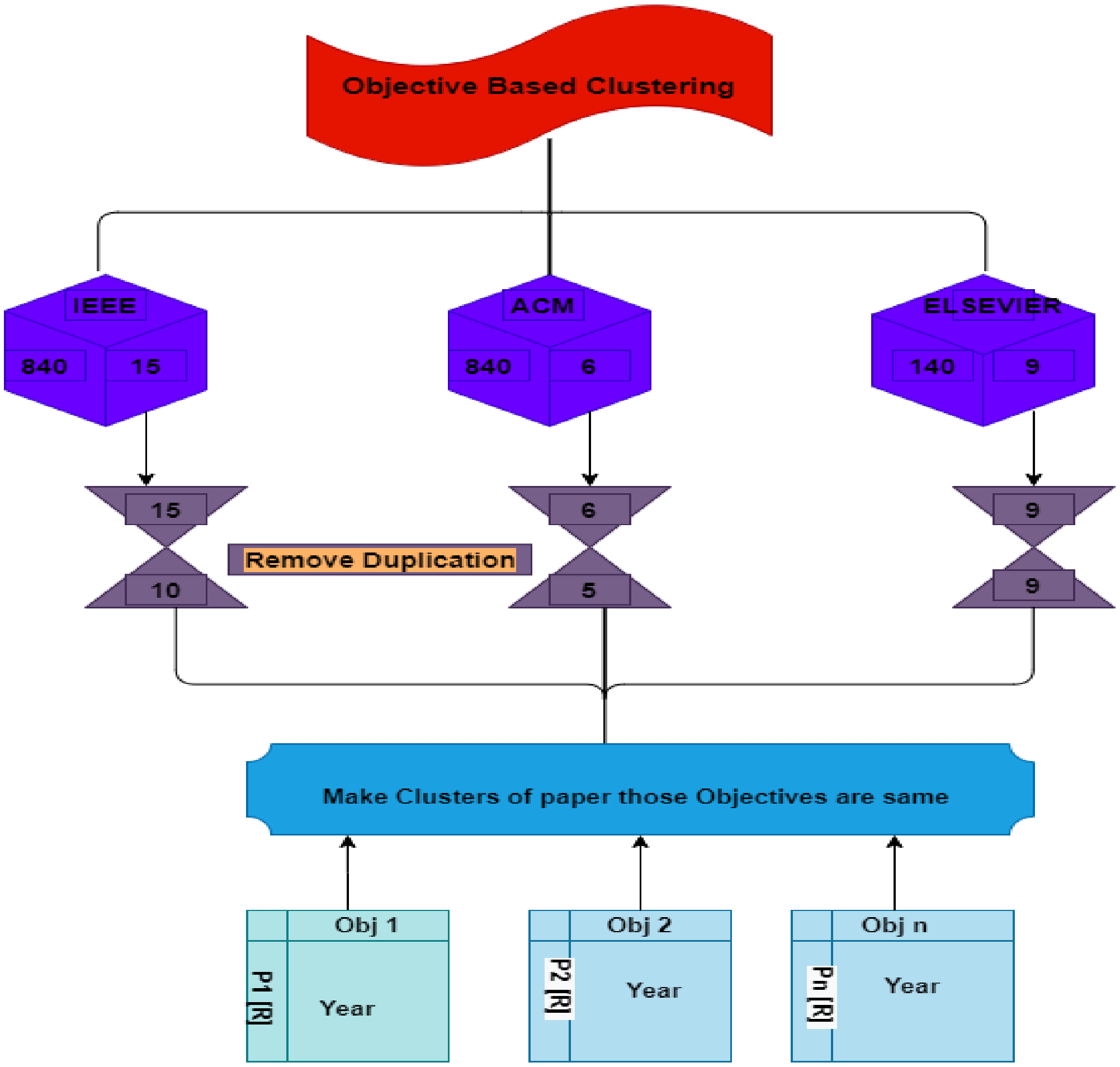

*Figure 9: Objective Based Clustering*

In figure 9, the objective-based clustering is discussed, read the objectives of all the research papers that come after the abstract filtering process and then make clustering of it. After that write the paper number, year, and the objective of multiple papers and check papers achieved their objectives or not in Table 2.

| **Paper** | **Year** | **Improved the Performance** | **Consistent and complete summaries compared to human written ones** | **Improve the quality of content selection** | **Covering the main content and reducing the redundant information** | **Minimizes the error rate** | **Improve efficiency** |
|---|---|---|---|---|---|---|---|
| **P1[2]** | **2019** | ☑ | | | | | |
| **P2[3]** | **2018** | ☑ | | | | | |
| **P3[4]** | **2018** | ☑ | | | | | |
| **P4[5]** | **2020** | ☑ | | | | | |
| **P5[6]** | **2018** | | ☑ | | | | |
| **P6[7]** | **2020** | | ☑ | | | | |
| **P7[8]** | **2018** | | | ☑ | | | |
| **P8[9]** | **2020** | | | ☑ | | | |
| **P9[10]** | **2019** | | | | ☑ | | |
| **P10[11]** | **2020** | | | | ☑ | | |
| **P11[12]** | **2020** | | | | ☑ | | |
| **P12[13]** | **2018** | | | | ☑ | | |
| **P13[14]** | **2019** | | | | ☑ | | |
| **P14[15]** | **2019** | | | | ☑ | | |
| **P15[16]** | **2020** | | | | | ☑ | |
| **P16[17]** | **2019** | | | | | | ☑ |
| **P17[18]** | **2016** | | | | ☑ | | |
| **P18[19]** | **2018** | ☑ | | | | | |
| **P19[20]** | **2020** | ☑ | | | | | |
| **P20[21]** | **2019** | | | ☑ | | | |

## 2.7 Quality Assessment:

Quality assessment (QA) criteria of selected study is designed by using the guidelines defined by the Kitchehem and Charters. QA is based on following quality questions (QQ):

Q1: Is the aim/objective of the study clearly stated?
Q2: Is study designed to achieve defined objectives?
Q3: Are findings and results clearly described in the study?
Q4: Is the study well referenced?

The QA of individual study is performed through quality points which are based on following rules:

Rule 1: If selected study answers the quality question, then it scores 1.
Rule 2: If selected study partially answers the quality questions, then its score is 0.5.
Rule 3: If selected study doesn't answer the quality question, then its score is 0.

The assessed quality of each of the selected study is given in Table 3 below. None of the study was excluded due to QA, since the QA score is ok for all studies.

**Table 3** Assessed quality of each selected study

| **PC** | **Q1** | **Q2** | **Q3** | **Q4** | **Total Points** | **Assessed Quality** |
|---|---|---|---|---|---|---|
| PC1 | 1 | 1 | 1 | 1 | 4 | Excellent |
| PC2 | 1 | 1 | 1 | 1 | 4 | Excellent |
| PC3 | 1 | 1 | 0 | 1 | 3 | Good |
| PC4 | 1 | 1 | 1 | 1 | 4 | Excellent |
| PC5 | 1 | 1 | 1 | 1 | 4 | Excellent |
| PC6 | 1 | 1 | 0.5 | 1 | 3.5 | Excellent |
| PC7 | 0.5 | 0.5 | 0.5 | 1 | 2.5 | Medium |
| PC8 | 1 | 0.5 | 0.5 | 1 | 3 | Good |
| PC9 | 1 | 1 | 1 | 1 | 4 | Excellent |
| PC10 | 1 | 1 | 1 | 1 | 4 | Excellent |
| PC11 | 1 | 1 | 0.5 | 1 | 3.5 | Excellent |
| PC12 | 1 | 1 | 1 | 1 | 4 | Excellent |
| PC13 | 1 | 1 | 1 | 1 | 4 | Excellent |
| PC14 | 1 | 1 | 1 | 1 | 4 | Excellent |
| PC15 | 1 | 1 | 0.5 | 1 | 3.5 | Excellent |
| PC16 | 1 | 1 | 1 | 1 | 1 | Excellent |

1. Low: If total points <=1
2. Medium: If total points=>1&<=2
3. Good: If total points=>2&<=3
4. Excellent: If total points=>3&=4

## 2.2 Detail Literature:

This section describes the detail literature review of the existing methods. There are three techniques which describe in this paper first is Extractive, second is abstractive and third is combine used of abstractive and extractive summarization technique. Extractive summarization generated the summary by adding the most relevant information in the sentence of the summary. Abstractive summarization understands the original text from the document and then creates new sentences from it [18] [57] [21].

### 2.2.1 Extractive Summarization:

That research [2] goal is to tackle the tough task of automatic text summarization of news content of the English language using classical statistical text processing principal. The objective of this paper is to acquire good content coverage with information variety and improved significant performance as compare to other summaries. The proposed system [2] is the improved version of the single document summarization. It performs all SDS tasks with some more subtask that minimizes the redundancy from the final summary. In past a lot of techniques are developed to overcome this problem like Artificial Bee Colony (ABC) (it is used for content coverage and redundancy removal) and the hybrid algorithm of fuzzy logic, cellular learning automata, Artificial Bee Colony (ABC), and PSO-GA, etc. are used for text summarization. Here cellular learning automata, Artificial Bee Colony (ABC) algorithm are used for calculating sentence scoring, and PSO-GA, fuzzy logic used for assigning the best weight to the extracted feature for final scoring. The proposed method of multi-document summaries is used to acquire content reportage with information variety and reduce the redundancy from the final summary. For the proposed method, input multi documents on which some preprocessing steps are required for example removing stop words, stemming, special character removal, segmentation, tokenization, etc. The next step is feature extraction as word feature, sentence feature and sentence scoring, and then summary generating process. Rule-based fuzzy logic is used for the final sentence scoring. Handling Redundant information is a big issue so cosine similar measure is used to eliminate sentences that have the same content to generate a final summary. For an experiment, DUC 2004 dataset is used for evaluation using Rough 2 and Rough 4. The result shows that the proposed MDS system improved the performance as compared to other methods and it gives high content coverage in the final summary.

The aim of paper [3] is to provide the main ideas in a document set in a short time. The objective of the research paper is to reduce the rebuilding error while creating a summary and improve the performance of the system. In this paper, the researcher developed a new technique, multi-document summarization using a distributed bag of word models. A distributed bag of words model is used for representing the sentences and documents. Firstly, train the PV-DBOW model to calculate the document vector for all documents in a set and then represent the key content of a document by the centroid vector. Then selects the summary sentence to decrease the reconstruction error while creating a summary. Beam search and sentence selection are used to reduce redundancy and improve the performance of the system. For the experiment DUC2006 and DUC2007 datasets were used. Results show that the proposed Methods achieve the best performance as compared to another unsupervised summarization system.

The aim of paper [4] is to address the problems of giving the information in compressed form to the reader. The objective of the paper is to improve the performance as compared to other systems and create an effective and human-readable summary. Finding out useful and relevant information related to a specific topic from the collection of documents is a big issue. To overcome this issue, the extractive summarization approach was developed. In this method, the blend of agglomerative hierarchical clustering and Latent Semantic Analysis (LSA) is used. Both approaches measure the semantic similarity between clauses and decrease dimensions by conserving only highly weighted vectors. LSA is an unsupervised tool. The latent Dirichlet Allocation model is used to recognize the significant term in the final summary. POS tagger is used to solve the problem of noise. For generating a summary, the selection algorithm term frequency-inverse document frequency (t f *IDF) is used. For the experiment DUC2004 dataset is used. Results show that the proposed Methods generate optimal machine summary which certifying high quality and efficiency.

The paper [7] aims are to convert long sentences into a few thematic short sentences. The objective of the paper is to provides us high-quality summarization. In this paper, we convert the original documents into phrase graphs, taking into account both linguistic and deep representations, then smear spectral clustering to obtain multiple clusters of sentences, and finally to form a final summary of each compressed cluster. In the Multi-sentence compression (MSC) method, each cluster that comprises a set of sentences is used to generate a single document summary. For the experiment, Multi-news and DUC-2004 datasets are used through the ROUGE metric to check the performance of the proposed method. The results show that the developed method produces high-quality summaries.

The Aim of the paper [8] is to extract a summary of documents. The objective of the paper is to increase the excellence of content variety in an extractive summarization system**.** We have two types two methods of extracting summaries. The ILP method is the simplest and effective way of extracting summaries. The main function picks a summary sentence that capitalizes on the score of each candidate. The main function contains a key function. Due to restriction summary should contain a topical world. The Greedy method ranks the sentence based on their score from the sentence scoring step. This helps us to extract the top-rank sentences with the highest scores in summary**.** For experiment DUC 2001 and DUC2002 dataset is used and the result shows high-quality summaries.

The paper [10] aims are to generate a summary that has these three feature content coverages, reduces redundancy, and used informative words. The objective of the paper is to facilitate users and to enter the information in brief form without missing any information. In this method, there are various steps. In the first step from many documents, we filter unnecessary documents and make one new document. In the second step, we weight various text feature which helps us identifying the related sentence in the document. The last summary has been generated by ordering the relevant sentence. Shark Smell Optimizer SSO is used to score the sentence. For the experiment, DUC 2004, DUC2006, DUC2007, TAC08, TAC11 dataset, and multilingual dataset MultiLing13 is used. The results show that the proposed model improves performance as compared to another summarizer.

The paper [11] aims to extract the most relevant and vital content from the specified length of the collection of documents and generate the best summaries. The objective of the paper is to improve performance and generate high-quality summaries. The data-driven text summarization method is proposed which works using differential evolution (DE). For the optimizing tasks, Differential evolution is distributed parallel on a grid. The high processing quantity contempt the difficulty of the linguistic method particularly when there are multiple documents. Parallel computing and HPC grid are used to obtain the appropriate accuracy and efficiency to run a summarization model. Higher performance computing is used to decrease the time of resolution of the assumed problem. For the experiment DUC2002 dataset is used. ROUGE metric is used for checking performance and quality. The results show that the anticipated method advances the excellence and performance of summarization.

The aim of the paper [12] is to generate a summary that contains these elements content coverage, minimize redundancy, and use a relevant sentence in summary. The objective of this paper is to advance the readability and comprehensiveness of the summary. The planned method is generic extractive multi-document summarization. In this method, a multi-objective optimizer is used. For the experiment, DUC2002 datasets are used. The results show that the proposed method shows the best performance as compared to other state of art systems.

The aim of the paper [13] is to generate a non-redundant and precise summary with vital information from the collection of documents. The objective of the paper is to extract relevant sentences and improve performance as compared to another existing system. The proposed method develops a concept-based integer linear programming (ILP) model for multi-document summarization. It integrates the centrality and place feature to extract the less related sentence and measure the significance of a sentence while generating a summary. The centrality-based method is used in the sentence clustering model. For experiment DUC 2001 to 2004 dataset is used which shows that the proposed method achieves high performance as compared to other existing methods.

The paper [14] aim is to reduce a similar sentence from the collection of document and then generate a concise text summary. The objective of the paper is to achieve high accuracy in the final summary result. To reduce the redundancy of the sentence the researcher proposed a method. The group of Text Rank and Maximal Marginal Relevance (MMR) is used in it. In this method, Text Rank is used to extract the important sentence whereas Maximal Marginal Relevance (MMR) is used to minimize the redundant sentences. For the experiment two well know online news articles Alexa Rank and Detik are used.MMR performs very well and shows good performance so no redundant sentence is found in the summary.

The aim of the paper [16] is to generate a short version of the document without losing any important content. The objective of the paper is to improve the performance, generate informative and readable summaries. The proposed method ExDos has used the benefits of both the supervised and unsupervised method. The developed method is the first approach that used both methods combine in a single frame for multi-document summarization. The proposed method is used to reduce the blunder rate of the classifier from each cluster using dynamic logic feature weight. This method is also used to measure the quantity of the significance of features in the multi-document summarization process. For the experiment DUC2002 and CNN/Daily Mail is used. The result shows the proposed method achieved significant performance as compared to existing systems.

The aim of the paper [17] is to brief the information from documents by keeping key points and reducing irrelevant information. The objective of this paper is to read the main idea of the document easily. This method has numerous steps. Firstly, all documents are separated into individual sentences then phrases are tokenized and stops will be removed. In the final process, the roots of the remaining phrases will be extracted by the porter streaming algorithm. In the end, the corresponding weight will be calculated for the whole document by the matrix data structure. In this method, a multi-object artificial bee colony algorithm is used and OpenMP (API) is used for parallel programming. Parallel multi-object bee colony work as real behavior of honey bee. For the experiment DUC2002 dataset is used. The proposed method improves performance, good quality summary, and provides a very fast speed up of 55.50 for 64 threads with effectiveness of 86.72%.

Qiang and Chen [18], develop a multi-document summarization method using a close pattern, which is also known as pattern-based summarization (Pat Sum). It is used for content coverage and removes redundant content from the documents. The researcher used the extractive, unsupervised, and generic multi-document summarization method. The pattern-based method takes multiple documents as input, using a closed sequential pattern method to represents the sentence, then sentences are ranked for selecting the most meaningful sentence, and for reducing redundancy maximal marginal relevance MMR algorithm is used to select the sentences which are not similar from the one already selects. In this paper, the Pat Sum method is compared with different previous methods like term and ontology-based system, etc. The proposed method used the benefits of both term and ontology-based methods and avoiding their weakness. The advantage of this system is, it captures more informative terms, reduces redundancy, and improves the performance of the system as compared to other methods. For the experiment DUC2004 dataset is used.

Rautray and Balabantaray et al [19], proposed a multi-document summarization method using a Cuckoo search, to find the accurate summary from the collection of documents. It creates a generic extractive summary. The step involved in this method are initial processing, input, and summary representation, and after that cuckoo search algorithm is used. In this paper, the proposed system is compared with the previous method like particle swarm-based and cat swarm Optimization-based summarizer systems, etc. to validate the non-redundancy, readability, and cohesiveness. The performance of the proposed approach is significantly improved as compared to another system. It also improves the readability of the summary. For the experiment DUC2006 and DUC2007 used the dataset.

Bidoki and Moosavi et al [20], develop a semantic approach of an unsupervised system for extractive multi-document summarization by the blend of statistical learning and graph-based template. It is a generic and language-free model. This method aims to consider 3 factors in generating summary information significant, high content coverage, and reduce repetition in the information. The method learns an iconic representation of phrases from the set of given documents through the word2vec model. This system used an expansion algorithm, graph-based estimation, conceptual density tunning, and clustering approach to enhance and increase the efficiency of the summarization method. For the experiment DUC2002 and DUC2006 are used. The suggested solution enhances the overall working of the system as comparing another system.

Mao and Yang et al [21], develop three methods for generating a single document summary by using the combination of supervised and unsupervised learning. The proposed three techniques are used for assessing the significance of sentences and also check the relation of sentences at the time. Fist technique is used for sentence scoring using two methods (supervised and graph model), the second method uses the score of an unsupervised method as the quality of supervised learning and the last technique uses the score of supervised learning as a previous value of nodes in the graph model. These three techniques improve the accuracy of sentence scoring in documents. For experiment DUC2001 and DUC2002 dataset is used to shows all three techniques have acquired a good result and better for extracting summary using supervised and unsupervised learning. It also validates that the precision of clause selection in a graph-based model is only achieved from previous knowledge.

Yadav et al [47], developed a method for selecting essential document opinionated sentences. The procedure is as follows: On a dataset, it first does stop word removal, sentence splitting, stemming method, and part-of-speech tagging. As a result, the method uses the SentiWordNet database to assign a sentiment score to each word based on its POS tag. Finally, the sentiment score of a sentence is equal to the total of the sentiment scores of all the words in it.

Srikanth et al [48], use BERT existing model to construct extractive summary by grouping the embeddings of sentences using Kmeans clustering, but to also present a dynamic method for determining the appropriate number of sentences to select from clusters. The project aims to provide higher-quality summaries by adding reference resolution and dynamically constructing summaries of appropriate sizes based on the text. For the experiment CNN/DailyMail dataset is used.

Rehmat et al [52], proposed the method RedunWSD method that used Word sense disambiguation method to remove redundancy from the collection of documents. The redundancy issue still not resolve yet. The disambiguating a word's sense helps in getting accurate semantic similarity score between two sentences. It takes into account the arrangement of words between two phrases in addition to the semantic similarity score. This study also has the benefit of recognizing repetitive sentences based on their meaning. The developed RedunWSD technique is based on the Lin semantic similarity measure and the word order similarity measure, both of which are sense specific.The semantic similarity metric is used to determine how similar two concepts are conceptually. WordNet is used to determine semantic similarity. Lin similarity is employed since it is a frequently used semantic similarity metric. For experiment, DUC-2005, DUC-2006, and DUC-2007 datasets is used to evaluate the result of proposed method with state of art methods.

### 2.2.2. Abstractive Summarization:

The aim of paper [5] to improve the content coverage to decrease the duplication problem in the generated summary. The objective of the research paper is to advance the excellence and readability of the created summary and also achieve higher performance and reduce redundancy. In this paper, the encoder and decoder method based on a double attention pointer network (DAPT) is developed. In the double attention pointer network, the self-attention method is used to collect important content from the encoder. The soft attention and pointer network are used to generate more comprehensible content. The improved method reduces the duplication problem and advances the quality of the summary. The combined method of scheduled sampling and reinforcement learning (RL) are used to produce new training methods to enhance the model. For the experiment two datasets are used, LCSTS and CNN/Daily Mail. The result shows that the developed method generates a more accurate and steadier summary and also improves the overall performance of the system as compared to another system.

The aim of the paper [9] is to generate a concise and comprehensive summary that contains important information in the input document. The objective of the paper is to improve the quality of the important sentence in the document by handling the issue of incomplete and duplicate sentences and generate high-quality summaries. In this model researcher used two techniques firstly combine a discriminative model D and a reproductive model G through adversarial learning. The generative model G takes the original document as input, performs sequence algorithm maximum likelihood estimation (MLE) on it, and generates a summary. The discriminator model D is the language model which is used to generate a summary by G avoiding the previous issue binary classifier discriminator. These two models are used to create high-quality summaries. To improve the importance of the informative sentence, generative model G used LSTM encoder and LSTM decoder to perform script categorization task and syntax annotation task. Multi-document summarization is classified into extractive and abstractive summarization. In extractive, the most relevant sentence is extracted to generate a summary whereas in the abstractive, reformulation of the document. The goal of an abstractive summary is to generate a precise and comprehensive summary of the collection of documents. The researcher proposed a plausibility-promoting reproductive adversarial network for abstractive summarization with multi-task constraints. This method is used to advance the performance of text summarization. This method is used to generate informative and grammatical error-free summaries. For the experiment, CNN/Daily Mail and Giga word are used. ROUGE is used for calculating performance. This method improves performance, generates high-quality, and informative summaries.

The aim of the paper [15] is to propose an abstractive text summarization system for multiple documents. The object of the paper is to generate an accurate summary that contains all the main contents of all documents. For handling the redundancy issue research develop a hybrid method that used clustering, word graphs, and neuronal networks. In clustering, all data from multiple documents is alienated into all clusters equally on basis of content. In a word graph, the shortest path detection minimizes the text. For generating abstractive text summarization, the sequence-to-sequence method, and supervised recurrent natural network (RNN) are used. For the experiment DUC2004 dataset is used and check performance by ROUGE and BLEU metrics.

### 2.2.3 Extractive and Abstractive Summarization:

The aim of paper [6] to generate a summary that is much closer to the human-written summary**.** The objective of the paper is to advance the execution of the system and generate more readable and natural summaries. The researcher proposed an actual joint framework for multi-document summarization. The joint framework includes an extractive and abstractive model which creates a summary that is much closer to the human-written summary. In this method, a sequence-to-sequence model is used in which we input multiple documents and generate its precise summary. In the extractive layer, a single-layer bidirectional LSTM encoder is used to select the sentence which improves the efficiency of the summary. After that neural attention model is used. For the experiment, the CNN/Daily Mail corpus dataset is used. The planned method achieved better performance and improve the readability of summaries.

Rong et al [51], proposed the technique that based on BERT, named T-BERTsum.It is a topic-aware abstractive and extractive summarization method. To direct the generation with the topic, the encoded latent topic representation is first matched with the embedded representation of BERT using the neural topic model (NTM). Second, the transformer network is used to learn long-term dependencies in order to jointly investigate topic inference and text summaries in an end-to-end way. Third, the extractive model is layered with long short-term memory (LSTM) network layers to record sequence timing information, and the effective information is further filtered on the abstractive model via a gated network. For the experiment, CNN/DailyNews and XSum dataset is used to evaluate the result of proposed method with state of art methods. In table 4, the detailed methodology of the scheme with dataset is discussed. Mostly researcher used DUC dataset and some used CNN/Daily Mail and MultiLing13 etc.

Table 4: The detail methodology of the scheme

| **Paper** | **Technique/ Methodology** | **Description** | **Dataset** |
|---|---|---|---|
| P [2] | Rule-based fuzzy logic and Cosine similarity measure | Input multi documents on which some preprocessing steps are required for example removing stop words, stemming, special character removal, segmentation, tokenization, etc.<br>The next step is feature extraction for example word feature, sentence feature, and sentence scoring, and then summary generating process. Rule-based fuzzy logic is used for the final sentence scoring.<br>Cosine similar measure is used to eliminate sentences that have the same content to generate a final summary. | DUC2004 |
| P [3] | Beam search, PV-DBOW | A distributed bag of words model is used for representing the sentences and documents.<br>Firstly, train the PV-DBOW model to calculate the document vector for all documents in a set and then represent the key content of a document by the centroid vector.<br>Then selects the summary sentence to decrease the reconstruction error while creating a summary. Beam search and sentence selection are used to reduce redundancy and improve the performance of the system | DUC2006, DUC 2007 |
| P [4] | Latent Dirichlet Allocation | The blend of agglomerative hierarchical clustering and Latent Semantic Analysis (LSA) is used. | DUC2004 |
| | (LDA), hierarchical agglomerative clustering, Latent Semantic Analysis | Both approaches measure the semantic similarity between clauses and decrease dimensions by conserving only highly weighted vectors.LSA is an unsupervised tool.<br>The latent Dirichlet Allocation model is used to recognize the significant term in the final summary.<br>POS tagger is used to solve the problem of noise. | |
| P [5] | Double attention pointer network (DAPT), reinforcement learning (RL), LSTM | In the DAPT, the self-attention method is used to collect important content from the encoder.<br>The soft attention and pointer network are used to generate more comprehensible content.<br>The improved method reduces the duplication problem and advances the quality of the summary. The combined method of scheduled sampling and reinforcement learning (RL) are used to produce new training methods to enhance the model. | CNN/Daily Mail, LCSTS |
| P [6] | Single-layer bidirectional LSTM, Neural attention model | A sequence-to-sequence model is used in which we input multiple documents and generate its precise summary.<br>In the extractive layer, a single-layer bidirectional LSTM encoder is used to select the sentence which improves the efficiency of the summary.<br>After that neural attention model is used. | CNN/Daily Mail corpus |
| P [7] | Multi-sentence compression (MSC) | Convert the original documents into phrase graphs, taking into account both linguistic and deep representations, then smear spectral clustering to obtain multiple clusters of sentences, and finally to form a final summary of each compressed cluster.<br>In the Multi-sentence compression (MSC) method, each cluster that comprises a set of sentences is used to generate a single document summary. | Multi-news, DUC-2004 |
| P [8] | ILP, Greedy algorithm | The ILP method is the simplest and effective way of extracting summaries.<br>The main function picks a summary sentence that capitalizes on the score of each candidate.<br>Due to restriction summary should contain a topical world.<br>The Greedy method ranks the sentence based on their score from the sentence scoring step. This helps us to extract the top-rank sentences with the highest scores in summary**.** | DUC2002 |
| P [9] | LSTM, maximum likelihood estimation (MLE), generative model, discriminator model | Firstly, combine a discriminative model D and a reproductive model G through adversarial learning. The generative model G takes the original document as input, performs sequence algorithm maximum likelihood estimation (MLE) on it, and generates a summary.<br>The discriminator model D is the language model which is used to generate a summary by G avoiding the previous issue binary classifier discriminator. | CNN/Daily Mail and Giga word |
| P [10] | Shark Smell Optimizer SSO | From the Collection of documents, we filter unnecessary documents and make one new document.<br>Then assign weight to various text features which helps us identifying the related sentence in the document.<br>Shark Smell Optimizer SSO is used to score the sentence. | DUC 2004, DUC2006, DUC2007, TAC08, TAC11, MultiLing13 |
| P [11] | differential evolution (DE), Higher performance | For the optimizing tasks, Differential evolution is distributed parallel on a grid. | DUC2002 |
| | computing (HPC) | The high processing quantity contempt the difficulty of the linguistic method particularly when there are multiple documents.<br>Parallel computing and HPC grid are used to obtain the appropriate accuracy and efficiency to run a summarization model.<br>Higher performance computing is used to decrease the time of resolution of the assumed problem. | |
| P [12] | Multi objective optimizer | The planned method is generic extractive multi-document summarization.<br>A multi-objective optimizer is used.<br>The content coverage is expressed by cosine similarity between each one of the summary sentences. | DUC2002 |
| P [13] | Integer linear programming (ILP), centrality base strategy | ILP integrates the centrality and place feature to extract the less related sentence and measure the significance of a sentence while generating a summary.<br>The centrality-based method is used in the sentence clustering model | DUC2001-2004 |
| P [14] | TextRank, Maximal Marginal Relevance | To reduce the redundancy of the sentence Text Rank and Maximal Marginal Relevance (MMR) is used.<br>TextRank is used to extract the important sentence.<br>Maximal Marginal Relevance (MMR) is sued to minimize the redundant sentences. | Alexa Rank and Detik |
| P [15] | Supervised recurrent natural network (RNN) | For handling the redundancy issue research develop a hybrid method that used clustering, word graphs, and neuronal networks.<br>In clustering, all data from multiple documents is alienated into all clusters equally on basis of content.<br>In a word graph, the shortest path detection minimizes the text.<br>For generating abstractive text summarization, sequence to sequence method, and supervised recurrent natural network (RNN) is used. | DUC2004 |
| P [16] | ExDos, Supervised, unsupervised model | The ExDos has used the benefits of both supervised and unsupervised method.<br>The developed method is the first approach that used both methods combine in a single frame for multi-document summarization.<br>This method is used to reduce the blunder rate of the classifier from each cluster using dynamic logic feature weight.<br>Basically, it is used to measure the quantity of the significance of features in the multi-document summarization process. | DUC2002 and CNN/Daily Mail |
| P [17] | Multi-object artificial bee colony algorithm, OpenMP (API) | Firstly, all documents are separated into individual sentences then phrases are tokenized and stops will be removed.<br>The roots of the remaining phrases will be extracted by the porter streaming algorithm.<br>In the end, the corresponding weight will be calculated for the whole document by the matrix data structure.<br>A multi-object artificial bee colony algorithm is used and OpenMP (API) is used for parallel programming.<br>Parallel multi-object bee colony work as real behavior of honey bee. | DUC2002 |
| P[18] | Pat Sum, Maximal marginal | The pattern-based method takes multiple documents as input, using a closed sequential pattern method to represents the sentence. | DUC2004 |
| | relevance MMR algorithm | Then sentences are ranked for selecting the most meaningful sentence, and<br>For reducing redundancy maximal marginal relevance MMR algorithm is used to select the sentences which are not similar from the one already selects. | |
| P[19] | Cuckoo search algorithm | The step involved in this method are initial processing, input, and summary representation, and after that cuckoo search algorithm is used. | DUC2006, DUC2007 |
| P[20] | Expansion algorithm, graph -based estimation, conceptual density tunning | The method learns an iconic representation of phrases from the set of given documents through the word2vec model.<br>This system used an expansion algorithm, graph-based estimation, conceptual density tunning, and clustering approach to enhance and increase the efficiency of the summarization method | DUC2002, DUC2006 |
| P [21] | Graph-based model supervised and unsupervised learning, Cosine Similarity | Fist technique is used for sentence scoring using two methods (supervised and graph model),<br>The second method uses the score of an unsupervised method as the quality of supervised learning and<br>The last technique uses the score of supervised learning as a previous value of nodes in the graph model.<br>These three techniques improve the accuracy of sentence scoring in documents. | DUC2001, DUC2002 |

In Table 5, the summary of critical analysis of schemes is presented. This table show the scheme type, technique, dataset, achievement and criticism.

**Table 5 Summary of critical analysis of schemes**

| S.no | Reference | Type | Techniques | Dataset | Achievements/Result | Criticism |
|---|---|---|---|---|---|---|
| **1** | Qian et al, (2021). | Extractive-Abstractive | T-BERTSum, LSTM, NTM | CNN/Daily Mail ,XSum dataset | Generate High quality summary with outstanding consistency. | It has limited processing power for long articles with multiple topics [16]. |
| **2** | Rahman et al. (2021, February | Extractive | RedunSWD, WSD | DUC 2005, 2006 and 2007 | Reduce the redundancy to get well content and rich informative summary. | The performances of WSD methods for all datasets are not uniform. Due to the presence of more ambiguity in some dataset. The performance is |
| | | | | | | quite low for all WSD systems [17]. |
| 3 | Li, Z., Peng, Z., Tang, S., Zhang, C., & Ma, H. (2020). | Abstractive | Double attention pointer network (DAPT), reinforcement learning (RL), LSTM | CNN/Daily Mail, LCSTS | advance the excellence and readability of the generated summary | It doesn't improve much for short text summary data-set [12]. |
| 4 | Zhao, J., Liu, M., Gao, L., Jin, Y., Du, L., Zhao, H., ... & Haffari, G. (2020, July). | Unsupervised | Multi-sentence compression (MSC) | Multi-news, DUC-2004 | Produce high-quality summaries. | - |
| 5 | Yang, M., Wang, X., Lu, Y., Lv, J., Shen, Y., & Li, C. (2020). | Abstractive | LSTM, maximum likelihood estimation (MLE), generative model, discriminator model | CNN/Daily Mail and Giga word | improve performance, generate high-quality, and informative summaries. | The limited the enhancement and improvement of prediction accuracy. It is a significant loss of information in the prediction process and prediction accuracy is difficult to improve [18]. |
| 6 | Ghodratnama, S., Beheshti, A., Zakershahrak, M., & Sobhanmanesh, F. (2020). | Extractive | ExDos, Supervised, unsupervised model | DUC2002 and CNN/Daily Mail | achieved significant performance as compared to existing systems. | - |
| 7 | Zamuda, A., & Lloret, E. (2020). | Extractive | differential evolution (DE),Higher performance computing(HPC) | DUC2002 | improve the quality and performance of summarization. | It can't use for abstractive summarization. it is also not applicable for summarization with word length less than the shortest sentence, because it selects whole sentences to generate a summary [11]. |
| 8 | Sanchez-Gomez, J. M., Vega-Rodríguez, M. A., & Perez, C. J. (2020). | Extractive | Multi-objective optimizer | DUC2002 | improve the readability and comprehensiveness of the summary | - |
| 9 | Bidoki, M., Moosavi, M. R., & Fakhrahmad, M. (2020). | Extractive | Expansion algorithm, graph - based estimation, conceptual density tunning and clustering approach | DUC2002, DUC2006 | Improve the performance and reduce the redundancy. (recall 0.58249,0.32625) | It does not review the summary readability, and usually ignores the position of the phrase when placing them in the summary [4]. |
| 10 | Gunawan, D., Harahap, S. H., & Rahmat, R. F. (2019, November). | Extractive | TextRank, Maximal Marginal Relevance | Alexa Rank and Detik | achieve high accuracy in the final summary result | Some irregular words and initials might cause low accuracy [ 13]. |
| 11 | Bhagchandani, G., Bodra, D., Gangan, A., & Mulla, N. (2019, May). | Abstractive | supervised recurrent natural network (RNN) | DUC2004 | to generate an accurate summary that contains all the main contents of all documents. | It Fall into a local minimum, slow convergence, and system stagnation [19]. |
| 12 | Sanchez-Gomez, J. M., Vega-Rodríguez, M. A., & Pérez, C. J. (2019). | Extractive | multi-object artificial bee colony algorithm, OpenMP (API) | DUC2002 | improve performance, good quality summary, and provide a high-speed efficiency. | The unsatisfactory operational ability makes it assemble to delay and, in some cases, it easily become trapped in the local optima [20]. |
| 13 | Verma, P., & Om, H. (2019). | Extractive | Shark Smell Optimizer SSO | DUC 2004, DUC2006, DUC2007, TAC08, TAC11, MultiLing13 | facilitate user and to access the information in brief form without missing any information. | - |
| 14 | Mao, X., Yang, H., Huang, S., Liu, Y., & Li, R. (2019). | Extractive | Graph based model, supervised and unsupervised learning, Cosine Similarity | DUC2001, DUC2002 | Improve the accuracy of sentence selection. (Rouge 1, Rouge 2) (0.44149,0.19485) | The threshold value affects the performance [21]. |
| **15** | Patel, D., Shah, S., & Chhinkaniwala, H. (2019). | Extractive | Rule-based fuzzy logic and cosine similarity measure | DUC2004 | Performance improvement and reduction of redundancy. | The conversion of fuzzy terms into clear actual values result in reduced accuracy [3]. |
| 16 | Mani, K., Verma, I., Meisheri, H., & Dey, L. (2018, December). | Extractive | Beam search, PV-DBOW | DUC2006, DUC 2007 | Improve the performance and minimize reconstruction error. | It is complex because multi-beam transmission increases beam alignment delay and greater power distribution to transmit a large beam, while lower power to transmit a narrow beam requires a wider dynamic power amplifier range. [22]. |
| 17 | Hafeez, R., Khan, S., Abbas, M. A., & Maqbool, F. (2018, October). | Extractive | Latent Dirichlet Allocation (LDA), hierarchical agglomerative clustering, Latent Semantic Analysis | DUC2004 | Improve the performance and create an effective and human-readable summary. | It lacks in using Numerous significant Words, it failed to Improve new Document immediately [23]. |
| 18 | Gui, M., Zhang, Z., Yang, Z., Gu, Y., & Xu, G. (2018, April). | Extractive, abstractive | single-layer bidirectional, LSTM, neural attention model | CNN/Daily Mail corpus | improve the performance of the system and generate more readable and natural summaries | The limited the enhancement and improvement of prediction accuracy. It is a significant loss of information in the prediction process and prediction accuracy is difficult to improve [25]. |
| 19 | Long, D. H., Nguyen, M. T., Bach, N. X., Nguyen, L. M., & Phuong, T. M. (2018, December). | Extractive | ILP, Greedy algorithm | DUC 2001, DUC2002 | increase the quality of content collection in an extractive summarization system | It has an issue of expensive build time [24]. And it can fail when applied to severely corrupted data [25]. |
| 20 | de Oliveira, H. T. A., Lins, R. D., Lima, R., Freitas, F., & Simske, S. J. (2018, October). | Extractive | integer linear programming (ILP), centrality-base strategy | DUC2001-2004 | extract relevant sentence and improve performance as compared to another existing system | It has an issue of expensive build time [24]. |
| 21 | Rautray, R., & Balabantaray, R. C. (2018). | Extractive, | Cuckoo search algorithm | DUC2006, DUC2007 | Improve the performance and readability (summary accuracy 0.99 and 0.9951) | The number of iterations require to find the best solution [26]. |
| 22 | Qiang, J. P., Chen, P., Ding, W., Xie, F., & Wu, X. (2016). | Extractive, | Pat Sum, Maximal marginal relevance MMR algorithm | DUC2004 | Improvement in performance (Recall 0.102 and 0.020) and reduce redundancy | It is hardly affected by minimum support.it also affect the performance of system [27]. |

## 3. Performance Analysis:

This section shows the performance of each research paper along with its objectives. This table also shows, which proposed method improved the objective of the paper.

In table 6, the results and achievements of research papers were discussed. The comparison shows that paper [2,3,4,5,19,20] improve the performance of summary, paper [6,7] generate a Consistent and complete summary compared to human written ones, paper [8,9,21] improve the content selection, paper [10,11,12,13,14,15,18] cover the main content and reduce the redundancy, paper [16] minimize the error rate and paper [17] improve the efficiency of summary.

Table 6 Results of papers

| | **Improved the Performance** | **Consistent and complete summaries compared to human written ones** | **Improve the quality of content selection** | **Covering the main content and reducing the redundant information** | **Minimizes the error rate** | **Improve efficiency** |
|---|---|---|---|---|---|---|
| **P2** | **95%** | **-** | **-** | **-** | | |
| **P3** | **42.68%** | | | | | |
| **P4** | **95%** | | | | | |
| **P5** | **40.26%** | | | | | |
| **P6** | | **36.38%** | | | | |
| **P7** | | **42.32%** | | | | |
| **P8** | | | **0.3503** | | | |
| **P9** | | | **42.15%** | | | |
| **P10** | | | | **0.470** | | |
| **P11** | | | | **0.31872** | | |
| **P12** | | | | **0.564** | | |
| **P13** | | | | **40,76** | | |
| **P14** | | | | **0.5103,0.4257** | | |
| **P15** | | | | **0.3383,0.2894** | | |
| **P16** | | | | | **52.5,42.1** | |
| **P17** | | | | | | **97.47%,96.20%** |
| **P18** | | | | **0.102 and 0.020** | | |
| **P19** | **0.99,0.9951** | | | | | |
| **P20** | **0.58249,0.32625** | | | | | |
| **P21** | | | **0.44149,0.19485** | | | |

.

In literature, there are a lot of techniques that were used to reduce redundancy and give an informative summary. However, all those techniques have some limitations. Because of these limitations, the system performance reduces, and the system doesn't perform well and gave an accurate summary. The detail of limitation is discussed in section 3.1.

## 3.1 Open Challenges:

Explore whether the scheme has considered the identified challenges or not.

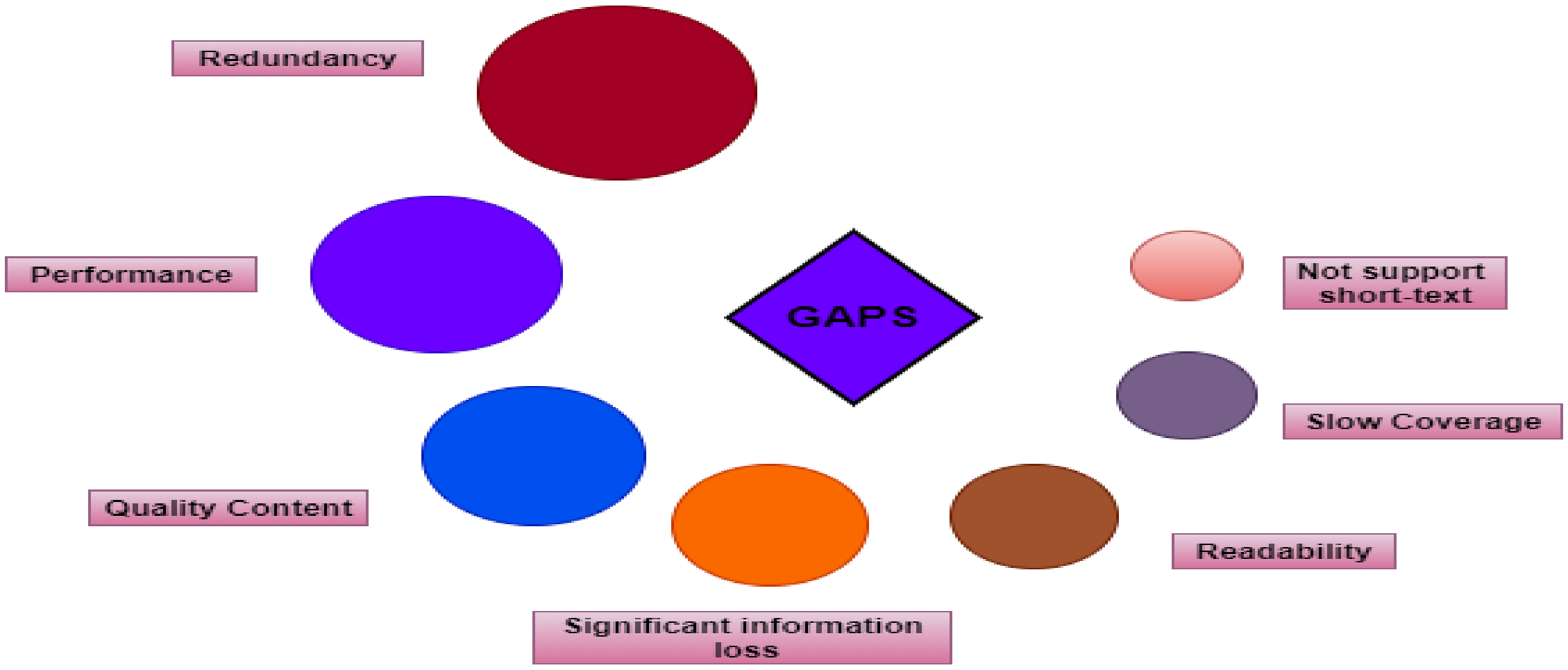

**Figure 10 Analysis of problems addressed by scheme listed in literature**

In figure 10, the main challenges that occur in multi-document summarization are discussed: Redundancy, Performance, Quality of content selection, Minimize error rate, Efficiency, slow coverage, Readability and Human written summaries

Papers use in this survey handle all these challenges and gave better results. Some papers reduce the redundancy from the summary, some papers improve the quality of content selection, some papers minimize the error rate and some generate a consistent and complete summary compared to human-written ones. In Table 7, the research challenge is discussed.

| Paper | Improve performance (Accuracy) | Consistent and complete summary compared to human written ones | Improve the quality of content selection | Reducing redundancy | Minimize Error rate | Improving efficiency |
|---|---|---|---|---|---|---|
| P [2] | ☑ | ☒ | ☒ | ☒ | ☒ | ☒ |
| P [3] | ☑ | ☒ | ☒ | ☒ | ☒ | ☒ |
| P [4] | ☑ | ☒ | ☒ | ☒ | ☒ | ☒ |
| P [5] | ☑ | ☒ | ☒ | ☒ | ☒ | ☒ |
| P [6] | ☒ | ☑ | ☒ | ☒ | ☒ | ☒ |
| P [7] | ☒ | ☑ | ☒ | ☒ | ☒ | ☒ |
| P [8] | ☒ | ☒ | ☑ | ☒ | ☒ | ☒ |
| P [9] | ☒ | ☒ | ☑ | ☒ | ☒ | ☒ |
| P [10] | ☒ | ☒ | ☒ | ☑ | ☒ | ☒ |
| P [11] | ☒ | ☒ | ☒ | ☑ | ☒ | ☒ |
| P [12] | ☒ | ☒ | ☒ | ☑ | ☒ | ☒ |
| P [13] | ☒ | ☒ | ☒ | ☑ | ☒ | ☒ |
| P [14] | ☒ | ☒ | ☒ | ☑ | ☒ | ☒ |
| P [15] | ☒ | ☒ | ☒ | ☑ | ☒ | ☒ |
| P [16] | ☒ | ☒ | ☒ | ☒ | ☑ | ☒ |
| P [17] | ☒ | ☒ | ☒ | ☒ | ☒ | ☑ |
| P [18] | ☒ | ☒ | ☒ | ☑ | ☒ | ☒ |
| P [19] | ☑ | ☒ | ☒ | ☒ | ☒ | ☒ |
| P [20] | ☑ | ☒ | ☒ | ☒ | ☒ | ☒ |
| P [21] | ☒ | ☒ | ☑ | ☒ | ☒ | ☒ |

In table 8, discuss the research gaps.

| Paper | Limitation/Gaps |
|---|---|
| P [2] | It produces incorrect similarity results toward vectors' magnitudes [18]. |
| P [3] | Beam search is complex because multibeam transmission increases beam alignment delay and greater power distribution to transmit a large beam, while lower power to transmit a narrow beam requires a wider dynamic power amplifier range [19]. |
| P [4] | LSA lacks in using Numerous significant Words, it failed to Improve the new Document immediately [19]. TFIDF is sometimes considered ad hoc [21]. |
| P [5] | This technique doesn't improve much for short text summary data-set [5]. |
| P [6] | LSTM method only considered prior context. So, the Identification of a word doesn't rely on the previous content but also on the following content [22]. |
| P [7] | - |
| P [8] | ILP has an issue of expensive build time [23].GA can fail when applied to severely corrupted data [24]. |
| P [9] | The LSTM method limited the enhancement and improvement of prediction accuracy. It is a significant loss of information in the prediction process and prediction accuracy is difficult to improve [25]. |
| P [10] | The SSO algorithm is a weakness in finding comprehensive solutions to some problems [26]. |
| P [11] | The challenge of DE is the unsteady convergence of the previous period, and the area is easy to fall optimal [27]. |
| P [12] | - |
| P [13] | ILP has an issue of expensive build time [23]. |
| P [14] | - |
| P [15] | RNN Fall into a local minimum, slow convergence, and system stagnation [28]. |
| P [16] | - |
| P [17] | The unsatisfactory operational ability makes it assemble to delay and, in some cases, it easily becomes trapped in the local optima [29]. |
| P [18] | Pat Sum is hardly affected by minimum support.it also affects the performance of the system [18]. |
| P [19] | The number of iterations to find the best solution [36]. |
| P [20] | It does not review the summary readability, and usually ignores the position of the phrase when placing them in the summary [20]. |
| P [21] | The threshold value affects the performance [21]. |

Patel and Shah et al [2] used a fuzzy logic algorithm. The disadvantage encountered when using fuzzy logic is that the conversion of fuzzy terms into clear actual values result in reduced accuracy[38].

Mani and Verma et al [3] used a beam search algorithm, the main disadvantage of a beam search algorithm is that they only focus on the next level When expanding the node, it can cause the node to be bad at the future level[40]. BS technique bounds the search space set aside only the best β node at every position to Extended loop. During a search, this strategy saves time. But the disadvantage is that the node Choice can ultimately eliminate talented candidates[41].

Hafeez and khan et al [4] used the LSA technique which is unable to relate a group of terms with underlining document theory[42]. LSA lacks in handling many important words, and it fails to improve the new document instantly[43]. The T- IDF technique is used in this paper which has some drawbacks like Since it is not acquired from any mathematical model of relevance analysis and interval distribution, it is sometimes considered ad hoc[44].

Zhao and Liu et al [7] used a Spectral clustering algorithm. This technique endures from dense similar matrix construction for huge content[45]. The Spectral clustering algorithm cannot deal with multiple categories of supervision and sometimes they show unstable effectiveness [46].

Sanchez-Gomez and Rodriguez [17] used an Artificial bee colony (ABC) model which has the drawbacks of poor exploitation and slow convergence[39]. The drawback of cosine similarity, the threshold value θ has affected the performance. If we set the threshold value too high then the purposed of minimizing or eliminating redundancy cannot be gained. If we set the threshold values too small then some important and important information may be eliminated[21].

Qiang and Chen et al [18] used the Pat Sum technique that has some drawbacks. User-specifies some restrictions (λ and sup) that can affect Pat Sum's accuracy. Here λ is a function that controls the tradeoff between the non-redundancy and context coverage. If we increase the value of λ, the accuracy will increase at the start and then at the end decreases. If the value of λ is small, the accuracy of Pat Sum is decreased dramatically because Pat Sum will not reduce too much information redundancy. It means Pat Sum is hardly affected by minimum support.

Rautray and Balabantaray et al [19] used a cuckoo search algorithm, the main disadvantage of the CS technique is the initial strategy. CS method random Initialize the position of the repeatable nest, then Increased the possibility of falling into local optimum solution[37]. The drawback of this algorithm is the number of iterations to find the Best solution[36].

The drawback of the paper[20], it does not consider the summary readability, and usually ignores the position of the clause when placing them in the summary. Structural similarity is not taken into account in the similarity's estimation module.

Mao and yang [21] used cosine similarity. The drawback of cosine similarity, the threshold value θ (redundancy)is affected the performance. If we set the threshold value too high then the purposed of minimizing redundancy cannot be gained. If we set the threshold values too small then a few important and useful data may be eliminated.

After discussing limitations, we conclude that there is still a need to develop a new method that overcomes these limitations by improving the overall performance of the system and generate a redundancy-free, informative summary. In our survey on multi-document summarization through a systematic literature review, our research integrates foundational insights from [60-76]. Moreover, [77-91] serve as pillars of knowledge, contributing to the foundation upon which our exploration and analysis are grounded

# 4. Conclusion:

Today, data extraction about a specific topic from the collection of documents is difficult. To overcome this problem multi-document summarization techniques are developed. These techniques handle content coverage, redundancy reduction, and relevant sentence extraction issues from multiple documents to generate a single document summary. There are a lot of techniques that were used to reduce redundancy and give an informative summary. However, all those techniques have some limitations. Because of these limitations, the system performance reduces, and the system doesn’t perform well and gave an accurate summary. The proposed method improved the accuracy and efficiency of the summary but the analysis has shown that higher accuracy with an effective algorithm is still needed in the field.

## 5. Future Work

Although the provided analysis and methodologies perform quite well many different improvements, adaptations, tests, and experiments are needed for future work. In the future, multi-document summarization may remove redundancy at a higher level and provide better results than the existing model. There is always a need for improvement in the work as no one achieves perfection.